\documentclass[12pt]{article}
\pdfoutput=1
\usepackage{amsmath,amssymb}

\usepackage{graphicx}% Include figure files
\usepackage{color}
\usepackage[bookmarksnumbered=true,colorlinks=true,linkcolor=black,citecolor=black]{hyperref}
\newlength{\extraspace}
\newlength{\extraspaces}
\def\bsklength{.8mm} %{2mm} % for more than double spacing
\newcommand{\beq}{\begin{equation}}
\newcommand{\eeq}{\end{equation}}

\newcommand{\bseq}{\addtocounter{subeqno}{1}\begin{subequations}}
\newcommand{\eseq}{\end{subequations}}
\font\mathscript=eusm10 at 12pt
\font\mathscripts=eusm7
\font\mathscriptss=eusm5
\newfam\mathscri
\textfont\mathscri=\mathscript
\scriptfont\mathscri=\mathscripts
\scriptscriptfont\mathscri=\mathscriptss
\def\mathscr#1{{\fam\mathscri\relax#1}}

\font\mathfrakt=eufm10 at 12pt
\font\mathfrakts=eufm7
\font\mathfraktss=eufm5
\newfam\mathfraki
\textfont\mathfraki=\mathfrakt
\scriptfont\mathfraki=\mathfrakts
\scriptscriptfont\mathfraki=\mathfraktss
\def\mathfrak#1{{\fam\mathfraki\relax#1}}

\def\CL{{\cal L}}

\def\CO{{\cal O}}

\renewcommand{\tilde}{\widetilde}

\renewcommand{\bar}{\overline}
\def\half{{\textstyle{1\over 2}}}
\def\pa{\partial}

\begin{document}
\setcounter{page}{0}
\addtolength{\baselineskip}{\bsklength}
\thispagestyle{empty}
\renewcommand{\thefootnote}{\fnsymbol{footnote}}        %for symbols

\begin{flushright}
arXiv:1209.1377 [hep-ph]\\
%{\sc DRAFT: \today}\\
\end{flushright}
\vspace{.4cm}

\begin{center}
{\Large
{\bf{Hidden Neutrino Gauge Symmetry}}}\\[1.2cm]
%{\large{\it{---??---}}}\\[1.2cm] %title
{\rm HoSeong La\footnote{hsla.avt@gmail.com}
}%          %author
\\[3mm]
{\it Department of Physics and Astronomy,\\[1mm]
Vanderbilt University,\\[1mm]              %address
Nashville, TN 37235, USA} \\[1.5cm]

\vfill
%{\sc Abstract}\\[1cm]
{\parbox{15cm}{
\addtolength{\baselineskip}{\bsklength}
\noindent
A new gauge force that acts exclusively on neutrinos is proposed. 
This new force violates neutrino flavors while masses are diagonal,
potentially opening a door for a new field theoretical treatment of
the neutrino oscillation.
The basic idea and a framework are presented. 

\bigskip
%Keywords: new gauge force, neutrino flavor violation, left-right symmetry\\
%PACS: 12.60.Cn, 14.60.Pq, 14.60.St, 11.30.Ly  
}
}

%{Submitted to {\it somewhere}}

\end{center}
\noindent
\vfill

%%%% table of contents %%%%
%\newpage
%\setcounter{page}{1}
%\pagenumbering{roman}
%\tableofcontents
%\vfill
%%%%%%%%%%%%%%%%%%%%%%%%%%%%%%%%%%%%%%%%

\newpage
\setcounter{page}{1}
\setcounter{section}{0}
\setcounter{equation}{0}
\setcounter{footnote}{0}
\renewcommand{\thefootnote}{\arabic{footnote}}  %for numbers
\newcounter{subeqno}
\setcounter{subeqno}{0}
\setlength{\parskip}{2mm}
\addtolength{\baselineskip}{\bsklength}

\pagenumbering{arabic}

%%%%%%%%%%%%%%%%%%%%%%%%%%%%%%%%%%%%%%%%

%\newsection{Introduction}
Nonzero neutrino masses clearly indicate the existence of physics beyond 
the Standard Model (SM). The oscillation data imply neutrino masses 
are of $\CO(1)$ eV or less and non-degenerate\cite{pdg}. 
There are various attempts of extending the SM to accommodate the current 
experimental data (see some recent reviews 
\cite{Altarelli:2011wd,Morisi:2012fg,Langacker:2011bi} among many others), 
but it is safe to say that none has been definitively successful due to 
lack of an explicit neutrino flavor violating 
structure\footnote{See \cite{Schechter:1981cv} for an earlier example of  
the neutrino flavor violation.}. 
So, there is a room for another proposal, which is drastically different 
from others.
In this Letter, we will briefly lay out the basic idea and a framework
to extend the SM with a new gauge force exclusively acting on neutrinos.
More details about the neutrino physics in this context
will be reported elsewhere in the future.

%\newsection{Basic Idea}
The basic ingredients we need for potentially successful neutrino physics,
which can explain the neutrino oscillation phenomenon field 
theoretically\cite{Giunti:1993se,Beuthe:2001rc,Akhmedov:2010ms,Wu:2010yr},
are that the new gauge force violates neutrino flavors and neutrino 
masses are non-degenerate\footnote{Even if one starts with tree-level
massless neutrinos, the degeneracy has to be broken to generate the neutrino 
oscillation\cite{Valle:1987gv}.}. 
We assume the new gauge force is abelian, U(1)$_\nu$, which is spontaneously 
broken at some energy scale above the electroweak (EW) scale. 
As a typical consequence, this spontaneous symmetry 
breaking  generates a mass for U(1)$_\nu$ gauge boson, but the pure massless 
neutrino sector still reveals the gauge symmetry. 
So, we can still take advantage of the gauge invariance in the pure neutrino 
sector. Then a flavor-violating Lagrangian for the pure neutrino sector can be 
constructed as
\beq
\label{e:p1}
\CL_\nu= -{1\over 4}F_{\mu\nu} F^{\mu\nu}+
i\sum_{i=1}^N
\bar{\psi_i}\gamma^\mu\left(\pa_\mu+ig_iA_\mu\right)\psi_i
-\sum_{i,j=1}^N\alpha_{ij}A_\mu\bar{\psi_i}\gamma^\mu \psi_j
-\sum_{i=1}^N m_i\bar{\psi_i}\psi_i,
\eeq
where $\alpha_{ij}=\alpha_{ji}$, $\alpha_{ii}=0$, and $\psi_i$ are neutrino 
mass eigenstates such that $m_i$ is the tree-level physical masses. 
$g_i$'s are preferably the same, but we reserve the possibility of different
$g_i$'s.
We assume that tree-level masses are generated by a mechanism outside
the pure neutrino sector such that they are free parameters here. 
Looking at this, one may hastily conclude that, even in the massless case,
this Lagrangian is not U(1)$_\nu$ gauge invariant because of the mixing
terms. However, being mass eigenstates does not guarantee they are also
U(1)$_\nu$ charge eigenstates so that one cannot just gauge transform
field variables in eq.(\ref{e:p1}).

In terms of a proper orthogonal transformation $\tilde{\psi}_i=O_{ij}\psi_j$,
where $O^TO=1$ for $O=(O_{ij})$, we can diagonalize the gauge coupling 
terms as
\beq
\label{e:8}
\CL_{\nu} = -{1\over 4}F_{\mu\nu} F^{\mu\nu}+
\sum_{i=1}^N i\bar{\tilde{\psi}_i}
\gamma^\mu\left(\pa_\mu+i\tilde{g}_i A_\mu\right)\tilde{\psi}_i
-\sum_{i,j=1}^N\tilde{m}_{ij}\bar{\tilde{\psi}_i}\tilde{\psi}_j,
\eeq
where $\tilde{\psi}_i$ are now U(1)$_\nu$ charge eigenstates and
\bseq
\begin{align}
\label{e:p3a}
(g_i\delta_{ij}+\alpha_{ij})&=O^T{\rm diag}(\tilde{g}_i)O, \\
\label{e:p3b}
{\rm diag}(m_i)&=O^T(\tilde{m}_{ij})O.
\end{align}
\eseq
The specific values of $O_{ij}$ can be easily derived from the above equations.

Eq.(\ref{e:8}) is manifestly gauge invariant if the mass term is diagonal.
However, since the Weak interaction violates U(1)$_\nu$ charge conservation
so that the mass generating mechanism can break U(1)$_\nu$ as well, 
the requirement of gauge invariance of mass term can be relaxed. 
Our intention is that, eq.(\ref{e:8}) has the gauge coupling terms diagonal 
in the charge eigenstates, while eq.(\ref{e:p1}) has mass terms diagonal in 
the mass eigenstates, but both cannot be diagonal at the same time.
Note that U(1)$_\nu$ charges are not quantized and different neutrinos
can carry different amount of charges.
Since neutrino flavor eigenstates and mass eigenstates are related by 
a unitary mixing matrix, $\tilde{\psi}_i$ are not necessarily the same
as the flavor eigenstates. 
So, eq.(\ref{e:p1}) implies neutrino flavor violation.

To demonstrate the idea, let us first examine the $N=2$ case.
Nevertheless, it should reveal some of characteristics for the three 
neutrino case. Solving eqs.(\ref{e:p3a}-\ref{e:p3b}), we can obtain
\beq
\label{e:p4}
\alpha_{12} ={\tilde{m}_{12}\over\tilde{m}}(\tilde{g}_1-\tilde{g}_2),
\eeq
where 
$\tilde{m} \equiv\sqrt{(\tilde{m}_{11}-\tilde{m}_{22})^2+4\tilde{m}_{12}^2}$.
This tells us that the flavor-violating coupling constants depend on the neutrino masses and that, to have flavor-violating interactions, 
the off-diagonal mass term for charge eigenstates must not vanish and 
different charge eigenstates must have different U(1)$_\nu$ charges. 
The diagonalized physical neutrino masses are given by
\bseq
\begin{align}
\label{e:p5a}
m_1 &=\half(\tilde{m}_{11}+\tilde{m}_{22}+\tilde{m}),\\
\label{e:p5b}
m_2 &=\half(\tilde{m}_{11}+\tilde{m}_{22}-\tilde{m}).
\end{align}
\eseq
This does not really tell us about any pattern of neutrino masses 
even if we assume off-diagonal mass is much smaller than the diagonal ones, 
and there is no other theoretical constraint we can impose 
(at least in the $N=2$ case).

%\newsection{Extending Gauge Symmetry}
However, if we extend the gauge symmetry, we can demand the gauge invariance 
of the off-diagonal mass terms under U(1)$_\nu$ to obtain an extra constraint.
Consider U(1)$_\nu\times {\rm U}(1)'={\rm U}(1)_1\times{\rm U}(1)_2$ 
for the pure neutrino sector , 
where ${\rm U}(1)'$, which is broken only by non-degenerate
neutrino masses, is a source of mixing. 

In terms of charge eigenstates under general U(1)$_1\times$U(1)$_2$, 
$\CL_{\nu}$ can be written in a manifestly gauge invariant form as, for $N=2$,
\beq
\label{e:3x}
\CL_{\nu} = -{1\over 4}\sum_{i=1}^N F^{(i)}_{\mu\nu} F^{(i)\mu\nu}+
i\sum_{i=1}^N\bar{\tilde{\psi}_i}
\gamma^\mu\left(\pa_\mu+i\tilde{g}_i A^{(i)}_\mu\right)\tilde{\psi}_i
-\sum_{i=1}^N\bar{\tilde{\psi}_i}m_{ij}\tilde{\psi}_j.
\eeq
Note that the mass terms are not diagonal and the off-diagonal parts
are not invariant under U(1)$_1\times$U(1)$_2$. But they are invariant
under the symmetric combination of ${\rm U}(1)_1\times{\rm U}(1)_2$, 
which we identify as U(1)$_\nu$ with a gauge field $A_\mu$.
Then 
\beq
\label{e:3y}
\left(
\begin{array}{cc}
gA_\mu & g'A'_\mu\\ g'A'_\mu &gA_\mu
\end{array}
\right)
\equiv O^T
\left(
\begin{array}{cc}
\tilde{g}_1A_\mu^{(1)} & 0\\ 0 &\tilde{g}_2 A_\mu^{(2)}
\end{array}
\right) 
O
\eeq
with
\beq
O={1\over \sqrt{2}}\left(
\begin{array}{cc}
1 & 1\\ 1 &-1
\end{array}
\right).
\eeq
This implies that $\tilde{m}_{11}=\tilde{m}_{22}$ for $\tilde{m}_{12}\neq 0$ 
to have diagonalized masses for mass eigenstates. 
Now eq.(\ref{e:3x}) becomes
\beq
\label{e:3z}
\CL_{\nu} =
-{1\over 4} F_{\mu\nu} F^{\mu\nu}-{1\over 4} F'_{\mu\nu} F'^{\mu\nu}
+i\sum_{i=1}^2\bar{{\psi}_i}
\gamma^\mu\left(\pa_\mu+ig A_\mu\right){\psi}_i
-gA'_\mu(\bar{\psi_1}\gamma^\mu\psi_2 +\bar{\psi_2}\gamma^\mu\psi_1)
-\sum_{i=1}^2 m_i\bar{{\psi}_i}{\psi}_i,
\eeq
where, from the gauge kinetic energy terms, we obtain $g=g'$ and
$\tilde{g}_1=\tilde{g}_2=\sqrt{2}g$.
Eq.(\ref{e:3z}) is invariant under U(1)$_\nu$, while
U(1)$'$ gauge field violates the flavor. Once the gauge symmetry
is extended, we can obtain flavor violating interaction even if 
$\tilde{g}_1=\tilde{g}_2$. In this case,
physical masses are $m_1=\tilde{m}_{11}+\tilde{m}_{12}$, 
$m_2=\tilde{m}_{11}-\tilde{m}_{12}$.
So, one can see that the compatibility of eq.(\ref{e:p1}) and eq.(\ref{e:8})
(or eq.(\ref{e:3x}) and eq.(\ref{e:3z}))
can provide nontrivial constraints on the properties of neutrinos.

%\newsection{Three Neutrinos}
In the $N=3$ case we further assume that the mixing between the first 
and third neutrinos are secondary so that it can be smaller than others, 
hence $\alpha_{13}$ should be smaller compared to others. But for the purpose
of explicit examples, we will assume $\alpha_{13}=0$. Otherwise, it becomes
rather cumbersome. We also assume $g_i$'s are the same.
Then from eq.(\ref{e:p3a}) we obtain
\beq
\label{e:10}
\tilde{O}={1\over\sqrt{2}}
\left(
\begin{array}{ccc}
\cos\xi & -\sqrt{2}\sin\xi& \cos\xi \\
-1 &0 & 1\\
\sin\xi & \sqrt{2}\cos\xi& \sin\xi
\end{array}
\right)
= O^T
\eeq
where the latter equality is up to an over-all sign,
$\tan\xi\equiv \alpha_{23}/\alpha_{12}$, and 
\beq
\tilde{g}_1=g-\sqrt{\alpha_{12}^2+\alpha_{23}^2}, \quad
\tilde{g}_2=g, \quad
\tilde{g}_3=g+\sqrt{\alpha_{12}^2+\alpha_{23}^2},
\eeq
such that
$\alpha_{12}^2+\alpha_{23}^2=(\tilde{g}_1-\tilde{g}_3)^2$.
Applying this to eqs.(\ref{e:p3a}-\ref{e:p3b}), we obtain
$\tan\xi= \tilde{m}_{23}/\tilde{m}_{12}=1$ and that $\tilde{m}_{ii}$'s must
be identical to be consistent, provided $\tilde{m}_{13}=0$.
The diagonal masses are given by 
\beq
\label{e:9}
m_1=\tilde{m}_{11} -\sqrt{\tilde{m}_{12}^2+\tilde{m}_{23}^2},\
m_2=\tilde{m}_{11},\
m_3=\tilde{m}_{11} +\sqrt{\tilde{m}_{12}^2+\tilde{m}_{23}^2}.
\eeq
such that
\beq
\label{e:9x}
m_2=\half(m_1+m_3).
\eeq
In this case, the flavor-violating prefactors are
\beq
\label{e:10a}
\alpha_{12}=\alpha_{23} 
={1\over 2\sqrt{2}}\left(\tilde{g}_3 -\tilde{g}_1\right)
\eeq
and coupling constants
\beq
\label{e:10b}
g_1=g_3 =\tilde{g}_2,\quad
g_2 =\half(\tilde{g}_1+\tilde{g}_3)=\tilde{g}_2,
\eeq
where we have imposed 
$\alpha_{13}=\tilde{g}_1\cos^2\xi+\tilde{g}_3\sin^2\xi-\tilde{g}_2\simeq 0$.

If we extend the gauge symmetry to U(1)$^3$, twisting gauge fields becomes
\beq
\label{e:11}
O^T
\left(
\begin{array}{ccc}
\tilde{g}_1A_\mu^{(1)} & 0 &0\\ 
0 &\tilde{g}_2 A_\mu^{(2)} &0\\
0 &0 &\tilde{g}_3 A_\mu^{(3)}
\end{array}
\right) 
O
=\left(
\begin{array}{ccc}
gA_\mu & g'A'_\mu & B_\mu\\ 
 g'A'_\mu &g'' A''_\mu & g'A'_\mu\\
B_\mu & g'A'_\mu &gA_\mu
\end{array}
\right),
\eeq
where
\bseq
\begin{align}
\label{e:14a}
gA_\mu &\equiv {1\over 4} \left(
\tilde{g}_1A_\mu^{(1)}+\tilde{g}_3 A_\mu^{(3)}
+2\tilde{g}_2 A_\mu^{(2)}\right),\\
g' A'_\mu &\equiv
{1\over 2}\left(\tilde{g}_1 A_\mu^{(1)}-\tilde{g}_3 A_\mu^{(3)}\right),\\
g'' A''_\mu &\equiv
{1\over 2}\left(\tilde{g}_1A_\mu^{(1)}+\tilde{g}_3 A_\mu^{(3)}\right),\\
\label{e:14d}
B_\mu &\equiv g'' A''_\mu - g A_\mu.
\end{align}
\eseq
In the above we have used the constraints due to the diagonalization of 
kinetic energy term such that $\tan\xi=1$, i.e. $\tilde{m}_{12}=\tilde{m}_{23}$,
and that $g=\tilde{g}_2$, $\tilde{g}_1=\tilde{g}_3$, $g'=\tilde{g}_1/2$,
and ${1\over g''^2}={1\over \tilde{g}_2^2}+{2\over \tilde{g}_1^2}$.
Since we would like to have all diagonal components of the rhs of 
eq.(\ref{e:11}) the same, which sets $g''A''_\mu=g A_\mu$, i.e.
$B_\mu=0$, equivalently $\alpha_{13}\simeq 0$. 
So, we can identify U(1)$_\nu$ with a gauge field given by eq.(\ref{e:14a}). 

Note that to make $\tilde{m}_{ij}\bar{\tilde{\psi}_i}\tilde{\psi}_j$,
$(i\neq j)$, gauge invariant under U(1)$^3$, U(1)$_\nu$ must be identified as
$A_\mu$ given by eq.(\ref{e:14a}) and for that $\tilde{m}_{ii}$'s must be the 
same. So, with one assumption $\tilde{m}_{13}=0$, identical $\tilde{m}_{ii}$'s
is necessary and sufficient condition for the flavor-violating lagrangian to
be fully gauge invariant even with the mass terms. In this sense,
$m_2=(m_1+m_3)/2$ is a non-trivial outcome.

In both case, this is possible if and only if
\beq
\label{e:15}
O=\half
\left(
\begin{array}{ccc}
1 & -\sqrt{2}& 1 \\
-\sqrt{2} &0 & \sqrt{2}\\
1 & \sqrt{2}& 1
\end{array}
\right)=\tilde{O}^T.
\eeq

\begin{figure}[t] %  figure placement: here, top, bottom, or page
\begin{minipage}[b]{0.32\linewidth}
\centering
\includegraphics[width=1.6in]{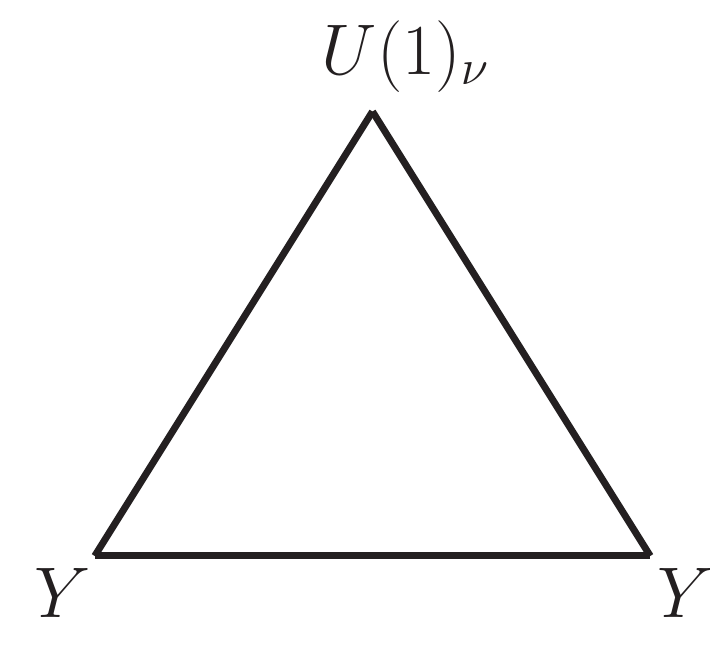} 
%\leavevmode\hbox{\epsfxsize=4truein\epsfbox{cho.eps}}
   \parbox{0.8\textwidth}{\vskip-6pt
   \caption{}
   \label{fig:1}}
\end{minipage}
%\hspace{0.1cm}
\begin{minipage}[b]{0.32\linewidth}
\centering
   \includegraphics[width=1.6in]{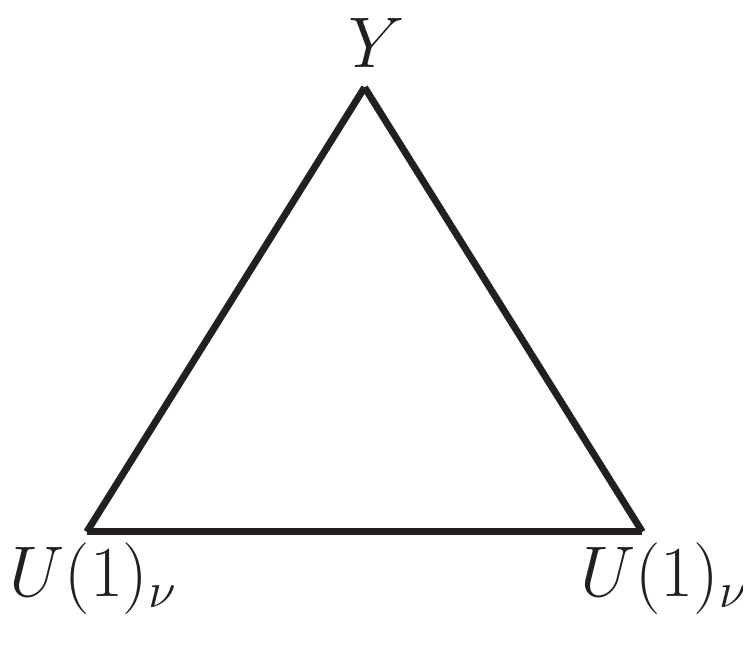} 
%\leavevmode\hbox{\epsfxsize=4truein\epsfbox{cho.eps}}
   \parbox{0.8\textwidth}{\vskip-6pt
   \caption{}
   \label{fig:2}}
\end{minipage}
\begin{minipage}[b]{0.32\linewidth}
\centering
   \includegraphics[width=1.6in]{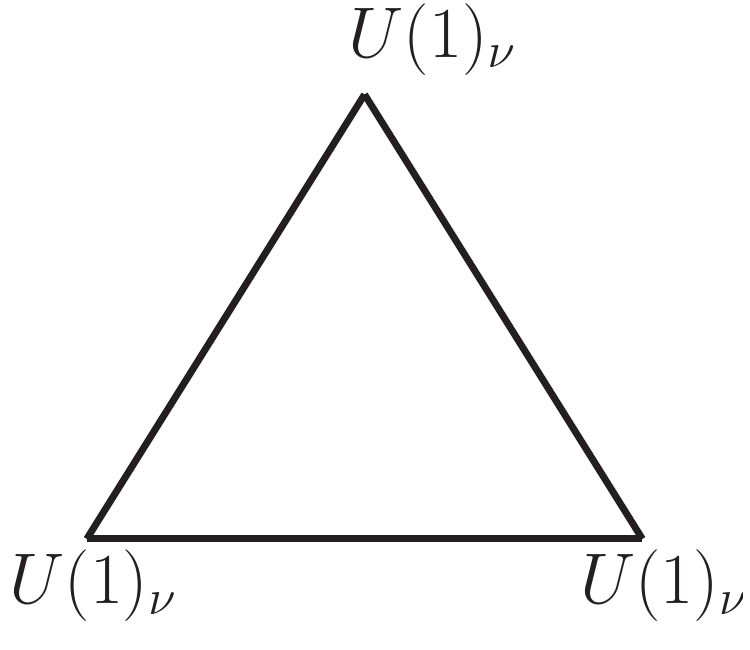} 
%\leavevmode\hbox{\epsfxsize=4truein\epsfbox{cho.eps}}
   \parbox{0.8\textwidth}{\vskip-6pt
   \caption{}
   \label{fig:3}}
\end{minipage}
\end{figure}

%\newsection{Anomaly Cancellation}
Once extra U(1)$_\nu$ is introduced, we need to worry about new anomalies.
We assume both chiralities of neutrinos carry U(1)$_\nu$ charges so that 
U(1)$_\nu$ is vector-like. 
The only new anomalies we need to worry about are those involving U(1)$_\nu$
as shown in Figs.\ref{fig:1}-\ref{fig:2}). 
(Fig.\ref{fig:3} vanishes because neutrinos are nonchiral under U(1)$_\nu$.) Fig.\ref{fig:2} cannot vanish in the SM, so U(1)$_\nu$ has to be a broken
symmetry in the SM energy scale and one needs to extend beyond the SM at the 
energy scale where U(1)$_\nu$ is unbroken. 
The cancellation of this anomaly requires left-handed neutrinos with 
opposite (hyper)charge, or the right-handed neutrinos need to carry 
(hyper)charges. The latter can be easily achieved by extending the SM 
to the left-right symmetric model\cite{PS,MP,Mohapatra:1979ia},
then all anomalies cancel regardless of $\alpha_{ij}$.
If the gauge symmetry is extended, since each U(1) acts on one generation of
leptons only, the same argument works.

Since the Weak interaction violates conservation of U(1)$_\nu$ charge,
this also indicates that
U(1)$_\nu$ must be broken at some scale above the EW scale.
Furthermore, for U(1)$_\nu$ to couple to neutrinos but not to charged 
leptons, we need a larger unbroken gauge group structure.
For example, the unbroken symmetry could be vector-like
SU(2)$'_V\times$U(1)$_X$ acting only on lepton doublets with $X$-charge $1/2$,
while quarks are neutral under them,
and U(1)$_\nu$ is the diagonal combination such that $Q_\nu=I_{V3}+X$.
The fact that SU(2)$'_V$ does not act on quarks distinguishes it from the 
vector part of SU(2)$_L\times$SU(2)$_R$.
U(1)$_\nu$ coupling to lepton doublets can be achieved in terms of an isospin 
projection operator $I^+\equiv \half(1+I_3)$ 
and U(1)$_\nu$ gauge boson
(or SU(2)$_{L,R}$ gauge bosons) transforms covariantly under SU(2)$_{L,R}$ 
(or U(1)$_\nu$, respectively) as vector matter\cite{La:2003jm}.
This property is inherited from the equivalent relationship between SU(2)$'_V$
and SU(2)$_{L,R}$ for SU(2)$'_V\times$SU(2)$_L\times$SU(2)$_R$ gauge symmetry.
U(1)$_\nu$ symmetry must be broken 
before the breaking of the left-right symmetry.
The neutrino masses generated by the seesaw mechanism in 
\cite{Mohapatra:1979ia} are non-degenerate and can be easily accommodated
into the framework presented here.

%\newsection{Results}
Although we have not performed detailed analysis of the system provided
for neutrino mixing and oscillations yet, there are rather desirable results.

In the example we presented for the three neutrino flavor case,
eq.(\ref{e:10}) with $\tan\xi=1$ and identical $\tilde{m}_{ii}$'s
are the compatibility conditions between eq.(\ref{e:p1}) with identical $g_i$'s 
and eq.(\ref{e:8}) (or eq.(\ref{e:3x}) and eq.(\ref{e:3z}) extended for 
three flavors) so that eq.(\ref{e:9}) is in some sense not entirely arbitrary.
The only assumption we make is $\tilde{m}_{13}=0$, then the equality of 
$\tilde{m}_{ii}$'s follows. 
Perhaps, a full analysis with $\tilde{m}_{13}\neq 0$ may 
reveal some incompatibility between eq.(\ref{e:p1}) and eq.(\ref{e:8}).
With extended gauge symmetry to U(1)$^3$, the equality of $\tilde{m}_{ii}$'s
is better clarified at the expense of having more symmetries.
Anyhow, if $m_2=(m_1+m_3)/2$ survives at least approximately after all 
higher order radiative corrections, neutrino masses can be estimated as 
$|m_1|\simeq 0.016$ eV, $|m_2|\simeq 0.018$ eV, and $|m_3|\simeq 0.051$ eV,
based on the current experimental data\cite{pdg}.
This is an interesting result, but we would rather not call it a prediction 
at this moment because the uniqueness is not clear and it will obviously
change in the case of $\alpha_{13}\neq 0$, although it is possible 
$\alpha_{13}$ could be negligible as indicated below eq.(\ref{e:14d}).

Our model as it is has an unfamiliar structure because U(1)$_\nu$ only acts on
neutrinos and the isospin doublet structure is not respected in the way
we assign U(1)$_\nu$ charges. 
Interestingly enough, U(1)$_\nu$ charge behaves
totally opposite to the electric charges as far as leptons are concerned.
To make it compatible we have to assume SU(2)$'_{\rm V}\times$U(1)$_X$, 
whose charged gauge bosons carry both U(1)$_\nu$ and U(1)$_{\rm EM}$ charges
(with the same sign) at low energy.
It will be interesting to ask if there is any direct way 
of checking the existence of U(1)$_\nu$. For example, charged leptons 
as well as the SM gauge bosons will have SU(2)$'_{\rm V}\times$U(1)$_X$ 
interactions at high energy well above the EW symmetry breaking.
For another example, a small directional variation of neutrino signals could indicate any deflection of neutrinos due to the emission of or scattering with 
U(1)$_\nu$ gauge bosons. The mass of U(1)$_\nu$ gauge boson is expected 
to be very light while the symmetry breaking occurs above the SM scale.
Nevertheless its coupling to neutrinos can be significant enough
such that the deflection of neutrinos from a faraway source
could be ``measurable'' in principle.
So, it will be worth to look into the possibility of unique processes due 
to U(1)$_\nu$.

We have not analyzed the consequence of neutrino flavor violating terms due
to nonvanishing $\alpha_{ij}$, but we expect they will play important roles 
in mixing and oscillation of neutrinos.
The examples we considered have $\alpha_{12}$ and $\alpha_{23}$, 
but no $\alpha_{13}$. So, the coupling between $\nu_1$ and $\nu_3$
will occur at higher order processes to make their effective couplings 
smaller than the other two. Since $\alpha_{ij}$'s can be 
different in principle, different amounts of neutrino species will come out 
from the same neutrino source and this process can be alternating to lead to 
the neutrino oscillation. In addition to neutrino masses, the whole process 
can be adjusted with parameters, $g, \alpha_{ij}$, 
which could account for the three degrees of freedom of mixing angles.

%\acknowledgement
I thank Tom Weiler for many illuminating conversations and for reading 
the manuscript. I also thank Stanley Deser and Jose Valle for helpful
communications.

\renewcommand{\Large}{\large}


\begin{thebibliography}{100}
\setlength{\itemsep}{-1.5mm}

\bibitem{pdg}
K.~Nakamura and S.T.~Petcov, in Particle Data Group,  
%``Review of Particle Physics (RPP),''
  Phys.\ Rev.\ D {\bf 86}, 010001 (2012)
(http://pdg.lbl.gov/2012/reviews/rpp2012-rev-neutrino-mixing.pdf)
and references therein.
  %%CITATION = PHRVA,D86,010001;%%

%\cite{Altarelli:2011wd}
\bibitem{Altarelli:2011wd} 
  G.~Altarelli,
  %``The Mystery of Neutrino Mixings,''
  arXiv:1111.6421 [hep-ph].
  %%CITATION = ARXIV:1111.6421;%%

%\cite{Morisi:2012fg}
\bibitem{Morisi:2012fg} 
  S.~Morisi and J.~W.~F.~Valle,
  %``Neutrino masses and mixing: a flavour symmetry roadmap,''
  arXiv:1206.6678 [hep-ph].
  %%CITATION = ARXIV:1206.6678;%%

%\cite{Langacker:2011bi}
\bibitem{Langacker:2011bi} 
  P.~Langacker,
  %``Neutrino Masses from the Top Down,''
  arXiv:1112.5992 [hep-ph].
  %%CITATION = ARXIV:1112.5992;%%

%\cite{Schechter:1981cv}
\bibitem{Schechter:1981cv} 
  J.~Schechter and J.~W.~F.~Valle,
  %``Neutrino Decay and Spontaneous Violation of Lepton Number,''
  Phys.\ Rev.\ D {\bf 25}, 774 (1982).
  %%CITATION = PHRVA,D25,774;%%

%\cite{Giunti:1993se}
\bibitem{Giunti:1993se} 
  C.~Giunti, C.~W.~Kim, J.~A.~Lee and U.~W.~Lee,
  %``On the treatment of neutrino oscillations without resort to weak eigenstates,''
  Phys.\ Rev.\ D {\bf 48}, 4310 (1993)
  [hep-ph/9305276].
  %%CITATION = HEP-PH/9305276;%%

%\cite{Beuthe:2001rc}
\bibitem{Beuthe:2001rc} 
  M.~Beuthe,
  %``Oscillations of neutrinos and mesons in quantum field theory,''
  Phys.\ Rept.\  {\bf 375}, 105 (2003)
  [hep-ph/0109119].
  %%CITATION = HEP-PH/0109119;%%

%\cite{Akhmedov:2010ms}
\bibitem{Akhmedov:2010ms} 
  E.~K.~Akhmedov and J.~Kopp,
  %``Neutrino oscillations: Quantum mechanics vs. quantum field theory,''
  JHEP {\bf 1004}, 008 (2010)
  [arXiv:1001.4815 [hep-ph]].
  %%CITATION = ARXIV:1001.4815;%%

%\cite{Wu:2010yr}
\bibitem{Wu:2010yr} 
  J.~Wu, J.~A.~Hutasoit, D.~Boyanovsky and R.~Holman,
  %``Neutrino Oscillations, Entanglement and Coherence: A Quantum Field theory Study in Real Time,''
  Int.\ J.\ Mod.\ Phys.\ A {\bf 26}, 5261 (2011)
  [arXiv:1002.2649 [hep-ph]].
  %%CITATION = ARXIV:1002.2649;%%

%\cite{Valle:1987gv}
\bibitem{Valle:1987gv} 
  J.~W.~F.~Valle,
  %``Resonant Oscillations Of Massless Neutrinos In Matter,''
  Phys.\ Lett.\ B {\bf 199}, 432 (1987).
  %%CITATION = PHLTA,B199,432;%%

\bibitem{PS}
J. Pati and A. Salam,
%``Lepton number as the fourth ``color'',''
Phys. Rev. D {\bf 10} (1974) 275.

\bibitem{MP}
R.N. Mohapatra and J.C. Pati, Phys. Rev. D {\bf 11} (1975) 566;
G. Senjanovi\'c and R.N. Mohapatra, Phys Rev. D {\bf 12} (1975) 1502.

%\cite{Mohapatra:1979ia}
\bibitem{Mohapatra:1979ia} 
  R.~N.~Mohapatra and G.~Senjanovic,
  %``Neutrino Mass and Spontaneous Parity Violation,''
  Phys.\ Rev.\ Lett.\  {\bf 44}, 912 (1980).
  %%CITATION = PRLTA,44,912;%%

%\cite{La:2003jm}
\bibitem{La:2003jm} 
  H.~S.~La,
  %``Implications of local chiral symmetry breaking,''
  hep-ph/0306223.
  %%CITATION = HEP-PH/0306223;%%

\end{thebibliography}
\end{document}